 \documentstyle[preprint,revtex]{aps}
 \tightenlines
 \begin{document}
 \draft
 \preprint{}
 \begin{title}
 \begin{center}
Exact Correlation Functions of Bethe Lattice Spin Models
in External Fields
 \end{center}
 \end{title}

 \author{Chin-Kun Hu\cite{huck} and N. Sh. Izmailian\cite{nick}}
 \begin{instit}
 Institute of Physics, Academia Sinica, Nankang, Taipei 
11529, Taiwan
 \end{instit}

 \date{\today}


\begin{abstract}
We develop a transfer matrix method to compute exactly the spin-spin 
correlation functions $<s_0s_n>$ of Bethe lattice spin models 
in the external magnetic field $h$ and for any temperature $T$.
We first compute  $<s_0s_n>$ for the most general spin - $S$ Ising  model,
which contains all possible single-ion and nearest-neighbor pair 
interactions. This general spin - $S$ Ising  model includes the
spin-1/2 simple Ising model and the Blume-Emery-Griffiths (BEG) model
as special cases. From  the spin-spin correlation functions,
we obtain functions of correlation length, ${\xi}(T,h)$,
for the simple Ising model and BEG model,
which show interesting scaling and divergent
behavior as $h \to 0$ and $T$ approaches the critical
temperature $T_c$.
Our method to compute exact  spin-spin correlation functions may
be applied to other Ising-type models on Bethe and Bethe-like
lattices.
\end{abstract}
\vskip 6 mm

\narrowtext

\begin{center}
{\bf I. INTRODUCTION}
\end{center}

The Bethe \cite{domb60} and the Bethe-like \cite{ef70} lattices 
have been widely used in solid state and statistical physics 
\cite{egg74,mz74,hu83,cm85,aai91,ef87,ty95,tm96,gu95,ku97,sms96,bbss96,smla96,mf94,ms93,mo96} 
as they represent underlying lattices for which many problems can 
be solved exactly. The Bethe lattice has attracted particular interest because 
it usually reflects essential features of systems, 
even when conventional mean - field theories fail \cite{gu95}. 

Besides the thermodynamic quantities, such as the magnetization, 
the specific heat, etc., the correlation function contains 
important information about a phase transition system 
\cite{stanley71} and is often studied by theoretical calculations 
\cite{avm96,wmtb76,dm95,ww76,mh75,yx94} and experimental 
measurements \cite{stinchcombe78,lr89a,lr89b,drort92}. It is 
widely believed that the singular behavior of physical quantities 
at the critical temperature $T_c$ of a second order phase 
transitions is related to the divergence of the correlation length 
$\xi$ at $T_c$ \cite{stanley71}.
In connection with this, the knowledge of the exact form of the 
spin-spin  correlation function is very crucial for locating phase 
transitions  and for analytical investigations of physical phenomenon. 
Furthermore, for many experiments, spin-spin correlation functions 
are most relevant, since they are measured by standard probes such 
as linear response to an adiabatic or isothermal applied field, 
or scattering of neutrons or electromagnetic waves 
\cite{stinchcombe78}. In the past several years, considerable 
progress has been achieved in the computation of the correlation 
function of statistical mechanical systems \cite{avm96,dm95}. 
However, the spin-spin correlation functions are exactly known 
only in a few models, including  the two-dimensional Ising model 
in zero magnetic field and at any temperatures \cite{wmtb76}, 
the planar Ising model in an magnetic field and  
exactly at $T_c$ \cite{dm95}.

A long-standing problem of statistical mechanics is the exact solution 
of the spin-spin correlation function for the Ising model in an 
external magnetic field and at any temperature. In this paper, we 
develop a transfer matrix method to compute exactly the spin-spin 
correlation functions $<s_ns_0>$ for the most general spin - $S$ model 
on the Bethe lattice for any temperatures $T$ and external field $h$.
The model include the Ising model \cite{baxter82} 
and the Blume-Emery-Griffiths (BEG) model \cite{beg71} as special 
cases. The correlation length ${\xi}(T,h)$ obtained from 
the spin-spin correlation function shows interesting scaling and 
divergent behavior as $h \to 0$ and $t \to 0$, where $t = (T - T_c)/T_c$ 
\cite{ih98}. Our results thus solve a long standing puzzle in the 
critical phenomena of the Bethe lattice Ising model. Our method may 
be applied easily to other Bethe lattice  spin models.

Although the free energy of the Cayley tree Ising model in zero 
external magnetic field is an analytic function of the 
temperature, the magnetization $m$ and the magnetic susceptibility 
$\chi$ of the central spin $s_0$ of the Bethe lattice Ising model 
have singular behaviors with the critical exponents $\beta$=1/2, 
$\delta$=3 and $\alpha$=0 \cite{baxter82}. However, there is no 
previous calculation which shows that $\xi$ of the Bethe lattice 
Ising model has singular behavior. Our work shows clearly that 
$\xi$ diverges at the critical point of $m$ and $\chi$ for the 
Bethe lattice Ising model.

The difference between the Cayley tree and the Bethe lattice has 
been discussed by Baxter \cite{baxter82}. In the Cayley tree, the 
surface plays a very important role because the sites on the 
surface is a finite fraction of the total sites even in the 
thermodynamic limit. As a consequences, the spin models on the 
Cayley tree exhibits quite unusual type of phase transition 
without long - range order 
\cite{ww76,mh75,yx94}; the calculated correlation functions do not 
show singular behavior \cite{ww76,mh75,yx94}. 
To overcome this problem one usually considers only properties of 
sites deep in the interior (away from the surface) of the Cayley 
tree. The union of such equivalent sites, with the same 
coordination number $q$, can be regarded as forming the Bethe 
lattice \cite{baxter82}. Thus the Bethe lattice is assumed to have 
translation symmetry like any regular lattices. 

In this paper, we demonstrate the crucial role of Bethe lattice
dimensionality in determining critical behavior of the correlation 
length and show clearly that correlation length $\xi$ diverges at 
the critical point of $m$ and $\chi$ for the Bethe lattice Ising 
model with critical exponent $\nu = 1$, which is different from the 
mean-field critical exponent $\nu = 1/2$ considered by Tsallis and  
Magalh\~es in a recent review paper \cite{tm96}, but is consistent 
with the critical exponent of the localization length associated with 
density-density correlator in the Bethe lattice Anderson model 
obtained by a supersymmetry method \cite{ef87}.
Our result gives independent support to the idea that mean - field 
approximation and Bethe lattice approach are not equivalent in 
principle \cite{gu95}.
In this paper we also analyze exact spin - spin correlation function
for the BEG model \cite{beg71} and find that at the tricritical point,
$\xi \sim t^{-1/2}$ with tricritical exponent ${\nu}_t = 1/2$.

Thus, we have solved a  long standing  puzzle in the critical 
phenomena of the Bethe lattice Ising model. Our approach  can be  
extended easily to other models (Potts, Ashkin-Teller, etc.) on 
the Bethe and Bethe-like lattices. In particular, our results for 
the Ising model on the Bethe lattice can be extended easily to 
Husimi lattices, because they can be related with each other in 
terms of the star - triangle transformation \cite{baxter82}. 

The outline of the paper is as follows: In Sec. II the most general 
spin - $S$  model is defined and the spin - spin correlation function 
of this model is evaluated in a closed form. 
In Sec. III we analyze the exact correlation function 
of the Bethe lattice spin - $1/2$ Ising model  and discuss 
the critical properties. In Sec. IV the critical behavior of the 
spin - spin correlation function is analyzed for the spin - 1 Ising 
model on the Bethe lattice. In Sec. V a brief discussion of our 
results is presented.

\begin{center}
{\bf II. A MOST GENERAL SPIN-$S$ MODEL}
\end{center}

The Ising - type models with spin greater than $1/2$ have  rich 
fixed-point structures. The great interest in these models arises 
partly from unusually rich phase transition behavior they display 
as their interaction parameters are varied, and partly from their 
many possible applications.

A spin-1 Ising model, which was initially introduced by 
Blume-Emery-Griffiths (BEG) \cite{beg71} in connection with phase 
separation and superfluid ordering in $He^3$ - $He^4$ mixtures. The 
BEG model has played an important role in the development of the 
theory of multicritical phenomena associated with various physical 
systems \cite{ls84} and has extensively been investigated 
in the literature  \cite{cm85,aai91}.
 
The spin-3/2 Ising model with dipolar and quadrupolar interactions 
was first introduced to explain phase transitions 
in $DyVO_4$ by Sivardiere and Blume \cite{sb72} and a different 
spin-3/2 Ising model for ethanol-water-carbon-dioxide was introduced
by Krinsky and Mukamel \cite{km75}. Another spin-3/2 Ising model was 
investigated by Sa Barreto and De Alcantra Bonfim \cite{bb91}.

Let us define the most general spin-$S$ model by the Hamiltonian 

\begin{equation}
\label{R1}
-{\beta}H=\sum_{<ij>}{H_1(s_i,s_j)}+\sum_i{H_2(s_i)},
\end{equation}
where $\beta = (k_BT)^{-1}$ and $s_i$ is a spin variable which takes 
a value on $\{-S,-S+1,...,S-1,S\}$. The first sum goes over all 
nearest - neighbor pairs of the Bethe lattice and the second over 
all sites. $H_1(s_i,s_j)$  contains all 
possible nearest - neighbor pair interactions and can be written as

\begin{equation}
H_1(s_i,s_j) = \sum_{\mu, \nu = 1}^{2S}J_{\mu \nu}s_i^{\mu}s_j^{\nu},
\label{R2}
\end{equation} 
$H_2(s_i)$  includes all possible single ion interactions
\begin{equation}
H_2(s_i) = \sum_{\mu = 1}^{2S}h_{\mu}s_i^{\mu}.
\label{R3}
\end{equation}
There are $S(2S+1)$ independent nearest - neighbor coupling constants 
$(J_{\mu\nu})$ in Eq. (\ref{R2}) and $2S$ external fields $h_{\mu}$ 
in Eq. (\ref{R3}). The Hamiltonian (\ref{R1}) can describe a variety 
of different model. The partition function has the form 

\begin{equation}
\label{R4}
Z=\sum_{\{s\}}\exp{\left\{\sum_{<ij>}{H_1(s_i,s_j)}+
\sum_i{H_2(s_i)}\right\}}.
\end{equation}

The advantage of the Bethe lattice is that for models formulated on it 
exact recursion relations can be derived. The calculation on a Bethe 
lattice is done recursively  \cite{baxter82}. When the Bethe lattice 
is ``cut'' apart at the central site 0, it separates into $q$ identical 
branches, each of which contains $(q-1)$ branches. Then the partition 
function of the model  can be written as

\begin{equation}
\label{R5}
Z_N=\sum_{{s}}\exp{(-\beta H)}=
\sum_{s_0}\exp{\left\{H_2(s_0)\right\}}g_N^{q}(s_0),
\end{equation}
where $s_0$ is a spin in the central site, $N$ is the number of 
generations ($N \to \infty$ corresponds to the thermodynamic limit), 
and $g_N(s_0)$ is in fact the partition 
function of one branch. Each branch, in turn, can be cut along any site 
of the 1th-generation which is nearest to the central site. The 
expression for $g_N(s_0)$ can therefore be written in the form 

\begin{equation}
g_{N}(s_0) = \sum_{s_1}\exp{\left \{H_1(s_0,s_1)+H_2(s_1) \right\}}
 g_{N-1}^{\gamma}(s_1),
\label{R6}
\end{equation}
where $\gamma = q-1$.

Consequently, we have $2S+1$ recursion relations for the 
$g_N(s_0)$,  where $s_0$ takes values $(-S, -S+1,....,S-1, S)$.
After dividing each of recursion relations on a recursion relation for
$g_N(S)$, we have $2S$ recursion relations for $x_N(s_0)$ 

\begin{equation}
\label{R7}
x_{N}(s_0)=\frac
{\sum_{s_1}\exp\left\{H_1(s_0,s_1)+H_2(s_1)\right\}x_{N-1}^{\gamma}(s_1)}
{\sum_{s_1}\exp\left\{H_1(S,s_1)+H_2(s_1)\right\}x_{N-1}^{\gamma}(s_1)},
\end{equation}
where 

\begin{equation}
\label{R8}
x_N(s_0)=g_N(s_0)/g_N(S)
\end{equation}  
and the equation for $g_N(S)$ is

\begin{equation}
g_{N}(S) = g_{N-1}^{\gamma}(S)
\sum_{s_1}\exp{\left \{H_1(S,s_1)+H_2(s_1) \right\}}x_{N-1}^{\gamma}(s_1).
\label{R9}
\end{equation} 

Since the right - hand side of Eq. (\ref{R7}) is bounded in $x_N$, 
it follows that $x_N$ is finite for $N \to \infty$. Through $x_N$, 
obtained by Eq.(\ref{R7}), one can express the density 
$m_{\mu}=<\left(\frac{s_0}{S}\right)^{\mu}>$ of central site 
(the symbol $<...>$ denotes the thermal average), where $\mu$ 
take values from $1$ to $2S$:
\begin{equation}
\label{R10}
m_{\mu}=<\left(\frac{s_0}{S}\right)^{\mu}>=\frac{1}{S^{\mu}}\frac
{\sum_{s_0}s_0^{\mu}\exp{\left\{H_2(s_0)\right\}x_N^{q}(s_0)}}
{\sum_{s_0}\exp{\left\{H_2(s_0)\right\}x_N^{q}(s_0)}},
\end{equation}
and other thermodynamic parameters. 

So we can say that the $x_N$ in the thermodynamic limit ($N \to 
\infty$) determine the states of the system. For this reason the 
recursion relations for $x_N$ given by Eq. (\ref{R7}) can 
be called the equations of state for the spin - $S$ model on the Bethe 
lattice. For example, at high temperatures the recursion 
Eq.(\ref{R7}) tends to a fixed point and therefore the system has an 
appointed magnetization $m$. 
Equations (\ref{R7}) and (\ref{R10}) are fundamental equations for the 
spin-$S$ model on the Bethe lattice.
 
For calculating the spin - spin correlation function, 
it is now convenient to write down the expression for partition 
function in the following form

\begin{equation}
Z_N = \sum_{{s_0s_1...s_n}}\exp{\left(\sum_{i=0}^{n-1}H_1(s_i,s_{i+1}) +
\sum_{i=0}^nH_2(s_i)\right)} g_N^{\gamma}(s_0)
g_{N-1}^{\gamma-1}(s_1) ...g_{N-n+1}^{\gamma-1}(s_{n-1})
g_{N-n}^{\gamma}(s_n) \label{R11},
\end{equation}
where $n$ denotes the number of steps from the central point 0.
Summing over $s_n, s_{n-1},..., s_1$ consistently, 
we obtain Eq.(\ref{R5}) again.
Then, the two-spin correlation function between $s_0$ and $s_n$, 
${\Gamma}(n) = \frac{1}{S^2}<s_0s_n>$, can be written as

\begin{equation}
{\Gamma}(n) = \frac{1}{S^2}\frac{\sum_{{s_0s_1...s_n}}s_0s_n
e^{\left(\sum_{i=0}^{n-1}H_1(s_i,s_{i+1})+\sum_{i=0}^nH_2(s_i)\right)}
x_N^{\gamma}(s_0)x_{N-1}^{\gamma-1}(s_1) ...x_{N-n+1}^{\gamma-1}(s_{n-1})
x_{N-n}^{\gamma}(s_n)}
{\sum_{{s_0s_1...s_n}}
e^{\left(\sum_{i=0}^{n-1}H_1(s_i,s_{i+1}) +\sum_{i=0}^nH_2(s_i)\right)}
x_N^{\gamma}(s_0)x_{N-1}^{\gamma-1}(s_1) ...x_{N-n+1}^{\gamma-1}(s_{n-1})
x_{N-n}^{\gamma}(s_n)}
 \label{R12}.
\end{equation}
We will show below that the calculation of $<s_0s_n>$ can be performed 
by a transfer - matrix method. Other techniques for the calculation of 
the spin - spin correlation function are not able to give the 
correlation in the presence of an external magnetic fields. 
For example, the results 
obtained in \cite{ww76,mh75,yx94} are appropriate only for $h$ strictly 
equal to zero without any symmetry breaking effects. 
It should be noted that in order to obtain results which are 
relevant for the Bethe lattice we must used the proper thermodynamic 
limit $N \to \infty$. 

The properties of the Bethe lattice can be investigateded by considering 
a Cayley tree with a very large number of generations ($N$) and one only 
looks at the thermal ensemble of the sites in the interior part of the 
first $n$ generations. Then the limit $N \to \infty$ is taken before 
$n \to \infty$.

We are interested in the case when the series of solutions of recursion 
relations given by Eq. (\ref{R7}) converges to a stable point as $N \to 
\infty$. 
In the thermodynamic limit ($N \to \infty$), we may expect $x_{N-n}$ 
does not depend on $n$, so that $x_{N-n}$ and $x_N$ can be regarded 
as the same fixed - point solutions $x$ of recursion relations given 
by Eq. (\ref{R7}), which corresponds to the behavior in the interior 
part of an infinite Cayley tree (i.e. the Bethe lattice). 
In this case 

$$
\lim_{N \to \infty}{x_{N-n}(s)}=x(s)
$$ 
for all finite $n$.
Then the recursion equations (equations of state) becomes 
\begin{equation}
\label{R13}
x(s_0)
=\frac
{\sum_{s_1}\exp\left\{H_1(s_0,s_1)+H_2(s_1)\right\}x^{\gamma}(s_1)}
{\sum_{s_1}\exp\left\{H_1(S,s_1)+H_2(s_1)\right\}x^{\gamma}(s_1)}
\end{equation}
and the spin-spin correlation function ${\Gamma}(n)$ given 
by Eq. (\ref{R12}) takes the form

\begin{equation}
{\Gamma}(n) =\frac{1}{S^2} \frac{\sum_{{s_0s_1...s_n}}s_0s_n
e^{\left(\sum_{i=0}^{n-1}H_1(s_i,s_{i+1})+\sum_{i=1}^nH_2(s_i)\right)}
x^{\gamma}(s_0)x^{\gamma-1}(s_1) ...x^{\gamma-1}(s_{n-1})x^{\gamma}(s_n)}
{\sum_{{s_0s_1...s_n}}
e^{\left(\sum_{i=0}^{n-1}H_1(s_i,s_{i+1}) +\sum_{i=1}^nH_2(s_i)\right)}
x^{\gamma}(s_0)x^{\gamma-1}(s_1)...x^{\gamma-1}(s_{n-1})x^{\gamma}(s_n)}
 \label{R14}.
\end{equation}

The correlation function of Eq. (\ref{R14}) can be expressed 
in the vector-matrix form. For this purpose, let us introduce 
a $(2S+1){\times}(2S+1)$ matrix ${\bf V}$ and a $(2S+1)$-component 
column vector $\bf R$. The elements of $\bf V$ are given by 
\begin{equation}
V_{ss'} = \exp{\left(H_1(s,s')+\frac{H_2(s)+H_2(s')}{2}\right)}
\left[x(s)x(s')\right]^{\frac{\gamma-1}{2}}, 
\label{R15}
\end{equation}
where $s$ and $s'$ independently take values $S,S-1,...,-S+1,-S$. 
The vector ${\bf R}$ and transposed vector ${\bf R}^T$ have elements

\begin{equation}
r_s = \exp{\left(\frac{H_2(s)}{2}\right)}x^{\frac{\gamma+1}{2}}(s)  
  \label{R16}.
\end{equation}
Let us also introduce the diagonal matrix  ${\bf S}$
$$
{\bf S} = \left(\begin{array}{cccc}
1&0&\ldots&0\\0&\frac{S-1}{S}&\ldots&0\\\vdots&\vdots&\ddots&\vdots
\\0&0&\ldots&-1\\ \end{array}\right).
$$
With these definitions we may rewrite Eq. (\ref{R14}) in the 
vector-matrix form

\begin{equation}
{\Gamma}(n) = \frac{{\bf R}^T{\bf S}{\bf V}^n{\bf S}{\bf R}}
{{\bf R}^T{\bf V}^n{\bf R}} 
\label{R17}.
\end{equation}

The transfer matrix {\bf V} is real-symmetric 
($V_{ss'}=V_{s's}$), and can be diagonalized by the transformation
$$
{\bf P}^{-1}{\bf V}{\bf P} = \left(\begin{array}{cccc}
{\lambda}_1&0&\ldots&0\\0&{\lambda}_2&\ldots&0\\\vdots&\vdots
&\ddots&\vdots\\0&0&\ldots&{\lambda}_{2S+1}\\ \end{array}\right),
$$
where ${\bf P}$ is a $(2S+1){\times}(2S+1)$ matrix with the elements 
$p_{ss'}$ and ${\lambda}_1$ , ${\lambda}_2$, ... and  
${\lambda}_{2S+1}$ - are the eigenvalues of the matrix ${\bf V}$. 
These eigenvalues can be obtained from the characteristic equation. 
Then the spin-spin correlation function of  Eq.(\ref{R17}) can be 
written as 
\begin{equation}
{\Gamma}(n) = \frac{{\bf R}^T{\bf S}{\bf P}{\bf L}{\bf P}^{-1}{\bf S}{\bf R}}
{{\bf R}^T{\bf P}{\bf L}{\bf P}^{-1}{\bf R}},
\label{R18}
\end{equation}
where ${\bf L}$ is a $(2S+1){\times}(2S+1)$ diagonal matrix
$$
{\bf L} = \left(\begin{array}{cccc}
{\lambda}_1^n&0&\ldots&0\\0&{\lambda}_2^n&\ldots
&0\\\vdots&\vdots&\ddots&\vdots\\0&0&\ldots
&{\lambda}_{2S+1}^n\\ \end{array}\right).
$$

After some algebraic manipulations, using  Eqs. (\ref{R13}), 
(\ref{R15}) and  (\ref{R16}), we may write the correlation function 
of  Eq.(\ref{R18}) as

\begin{equation}
{\Gamma}(n) = m_1^2 + \sum_{k=1}^{2S}A_kl_k^n 
\label{R19}.
\end{equation}
where $l_k \equiv {\lambda}_{k+1}/{\lambda}_1$. An explicit 
expression for $A_k$ has been given in Appendix. It should be noted 
that the obtained exact expression for the spin - spin correlation 
function depends on the ratios of the eigenvalues 
${\lambda_k}/{\lambda_1}$.
Thus, Eq. (\ref{R19}) together with Eqs. (\ref{R10}) and 
(\ref{R13}) give us full set of equations for investigation spin-$S$ 
model on the Bethe lattice.
In the following we turn to various examples. 

\begin{center}
{\bf III. SPIN-1/2 ISING MODEL}
\end{center}

We first consider a spin-1/2 Ising Hamiltonian at a temperature $T$ 
and an external magnetic field $h$
 
\begin{equation}
-\beta H = 4J\sum_{<ij>}s_is_j + 2h\sum_is_i 
\label{R20},
\end{equation} 
where $s_i=$ $+1/2$ or $-1/2$ and first term describes the 
ferromagnetic coupling $(4J)$ between the spin at site $i$ and $j$. 

For the magnetization ($m_1 = 2<s_0>$) of the spin in the central 
site we can obtain from Eq. (\ref{R10}) the following expression

\begin{equation}
m_1 = \frac{\exp{(2h)} -  x^{q}}{\exp{(2h)} + x^{q}} 
\label{R21},
\end{equation}
where $x$ is the fixed point of the recursion relations (\ref{R13}) 
in the thermodynamic limit

\begin{equation}
x^{q-1}\exp{(-2h)} = \frac{x\exp{(2J) - 1}}{\exp{(2J) - x}}
\label{R22}.
\end{equation}

The two-spin correlation function for the spin- 1/2 Ising model 
can be obtained from Eq. (\ref{R19})
 
\begin{equation}
{\Gamma}(n) = m_1^2 + A_1\left(\frac{{\lambda}_2}{{\lambda}_1}\right)^n 
\label{R23},
\end{equation}
with $A_1=1-m_1^2$ (see Appendix)  
and ${\lambda}_1$, ${\lambda}_2$ - are the eigenvalues of the 
$2{\times}2$ matrix ${\bf V}$. 

The elements of $\bf V$ are given by 
\begin{equation}
V_{ss'} = \exp{\left(4Jss'+hs+hs'\right)}
\left[x(s)x(s')\right]^{\frac{\gamma-1}{2}} 
\label{R24},
\end{equation}
where $s$ and $s'$ independently take values $\pm 1$ 
$$
{\bf V} = \left(\begin{array}{cc}V_{++}&V_{+-}
\\V_{-+}&V_{--}\\ \end{array}\right).
$$
The eigenvalues of the matrix  $\bf V$ can be obtained from the 
characteristic equation
\begin{equation}
{\lambda}^2 - {\lambda}(V_{++}+V_{--}) + V_{++}V_{--}-V^2 = 0.  
\label{R25}
\end{equation}
Using Eqs.(\ref{R22}) and (\ref{R24}) we find that
\begin{eqnarray}
{\lambda}_1 &=& 2\sinh{(2J)}\frac{\exp{(J+h)}}{(\exp{(2J)}-x)},
\nonumber \\
{\lambda}_2 &=& (2\cosh{(2J)}-x-x^{-1})\frac{\exp{(J+h)}}{(\exp{(2J)}-x)}  
\label{R26}.
\end{eqnarray}

Thus, the spin-spin correlation function can be expressed exactly as  
\begin{equation}
{\Gamma}(n) = m_1^2 +(1-m_1^2){\lambda}^n
\label{R27},
\end{equation}
where 

\begin{equation}
{\lambda} =\frac{{\lambda_2}}{{\lambda}_1}=  
\frac{2\cosh{(2J)}-x-x^{-1}}{2\sinh{(2J)}},
\label{R28}
\end{equation}
$m_1$ is the magnetization of the spin in the central site given by Eq. 
(\ref{R21}) and $x$ is the solution of the recursion relation given 
by Eq.(\ref{R22}). 

It is well known that the Ising model on the Bethe lattice 
exhibits ferromagnetism, with a critical point at $h = 0$, 
$x = 1$, $T = T_c$, where $J_c = \frac{1}{2}\ln{\frac{q}{q-2}}$ 
and critical exponents $\beta$,   $\delta$ and $\alpha$ have 
the ``classical'' values $\beta = \frac{1}{2}$, $\delta = 3$ 
and $\alpha = 0$ \cite{baxter82}. Let us now consider the 
general behavior of the spin-spin correlation function in 
the critical region.

First consider the case $h=0$ and $T=T_c$. From Eq.(\ref{R27}), 
we obtain ${\Gamma}(n) = (q-1)^{-n}$. Let us consider a Bethe 
lattice with coordination number $q$, whose dimension is defined 
by $d_n=\frac{\ln{C_n}}{\ln{n}}$ which tends to infinity with 
$n \to \infty$ for $q>2$ and equal 1 for $q=2$, where 
$C_n = \frac{q(q-1)^n-2}{q-2}$ is the total number of sites. 
We should note that for $q=2$ the Bethe lattice becomes the 
ordinary one-dimensional chain.   In the limit of large $n$, 
$d_n$ for all $q>2$ becomes
$$
d = \frac{n}{\ln{n}}\ln{(q-1)}.
$$
Thus we can write, for large $n$,  
\begin{equation}
{\Gamma}(n) = (q-1)^{-n} = n^{-d}.
\label{R29}
\end{equation}

Near the critical point, setting as usual 
$t = \frac{T - T_c}{T_c}$, we find that the spin-spin 
correlation function is

\begin{equation}
{\Gamma}(n) = \frac{\exp{\left(-\frac{n}{\xi}\right)}}{n^{d}}
\label{R30},
\end{equation}
where the correlation length $\xi$ is given by 
\begin{eqnarray}
\xi = \left[\ln{\frac{1}{(q-1){\lambda}}}\right]^{-1} 
 &=& \left[\ln{\left(\frac{1}{q-1}
    \coth{\frac{J_c}{1+t}}\right)} \right]^{-1}  \nonumber \\
 &{\sim}& \frac{q-1}{q(q-2)J_c}t^{-1}. 
\label{R31}
\end{eqnarray}

Thus, we find that the correlation length $\xi$ increases 
as the critical point is approached according to 
$\xi \sim t^{-\nu}$, with critical exponent $\nu = 1$. It 
is interesting to note that the correlation length $\xi$ shows 
interesting scaling and singular behavior near the critical 
point. While the Ising model on the Bethe lattice exhibits 
in general mean-field like phase transition with ``classical'' 
exponents, the critical behavior of the correlation length 
near the transition point coincides with the correlation 
length behavior in one-dimensional chain with critical 
exponent $\nu = 1$, which differs from its ``classical value'' 
$\nu = 1/2$ \cite{tm96}. The similar behavior of the localization length 
associated with density-density correlator can be observed 
in the Anderson model on the Bethe lattice \cite{ef87}.

If $h$ and $t$ are both sufficiently small in the critical 
region, then based on Eqs. (\ref{R22}), (\ref{R28}) and 
(\ref{R31}) the general behavior of the correlation length 
$\xi$ should be described by a scaling function $F$

\begin{equation}
\xi = t^{-1}F(ht^{-\frac{3}{2}})
\label{R32},
\end{equation}
where $F(x) = \left(f_1 + f_2x^{\frac{2}{3}}\right)^{-1}$ with 
$$
f_1=\frac{q(q-2)}{2(q-1)}\ln{\frac{q}{q-2}} \quad \mbox{and} 
\quad f_2^3 = 9\frac{q(q-2)}{(q-1)^2}.
$$
For a small $h > 0$ and $t \to 0$, we obtain from 
Eq. (\ref{R32}) the result:
\begin{equation}
\xi = f_2^{-1}h^{-2/3} \label{R33},
\end{equation}
i.e. the critical exponent is 2/3.

The bulk susceptibility per lattice site $\chi$ or the linear 
response against the field is derived from Eq. (\ref{R21}) as 
$\chi = \partial{m_1}/{\partial{h}}$
\begin{eqnarray}
\chi &=& \frac{\partial{m_1}}{\partial{h}} = {\chi}_0\left[1-
\frac{(q-1)(2\cosh{(2J)} - x - x^{-1})} {2\sinh{(2J)}}\right]^{-1}
\nonumber \\  
&=& \frac{{\chi}_0}{1-(q-1){\lambda}},
\label{R34}
\end{eqnarray}
where ${\chi}_0$ is nonsingular part of the magnetic 
susceptibility and given by

\begin{eqnarray}
{\chi}_0 &=& 2e^{-2J}\frac{(2\cosh{(2J)}-x-x^{-1})(2\exp{(2J)}-x-x^{-1})}
{\sinh{(2J)}(x+x^{-1}-2\exp{(-2J)})^2}
\nonumber \\ 
&=&(1-m_1^2)(1+{\lambda}).
\label{R35}
\end{eqnarray}
Thus, the  magnetic susceptibility $\chi$ can be written 
finally in the simple form
\begin{equation}
\chi =\frac{(1-m_1^2)(1+{\lambda})}{1-(q-1){\lambda}}.
\label{R36}
\end{equation}

By means of the fluctuation relation, $\chi = \sum {(\Gamma{(n)}-m_1^2)}$,
we can recover Eq. (\ref{R36}) through Eq. (\ref{R27}).
To proved this statement let us consider in more detail 
the fluctuation relation

\begin{equation}
\chi =\lim_{N \to \infty} \frac{1}{N_s}\sum_{ij}\frac{1}{S^2}
(<s_is_j>-<s_i><s_j>),
\label{fluc1}
\end{equation}
where the sum goes over all pairs of sites on the Bethe 
lattice and $N_s$ is the total number of sites. To carry 
out the summations in Eq. (\ref{fluc1}), we first  
note that by definition all sites on the Bethe lattice 
are equivalent and consequently 
$$
<s_i>=<s_0>,
$$
$$  
\sum_{ij}<s_is_j> = N_s \sum_{j}<s_0s_j>.
$$
Using these relations we may rewrite  Eq. (\ref{fluc1}) as
\begin{eqnarray}
\chi &=& \lim_{N \to \infty} \sum_{j}\frac{1}{S^2}(<s_0s_j>-<s_0>^2)
\nonumber \\  
&=&\lim_{n \to \infty} \sum_{j=1}^nc_j[\Gamma (j)-m_1^2],
\label{fluc3}
\end{eqnarray}
where $c_j$ is the number of sites which is $j$ steps away 
from the central site 0
\begin{equation}
c_j=q(q-1)^{j-1}, \quad \quad \quad  (c_0=1),
\label{fluc4}
\end{equation}
$\Gamma (j)$ is the spin - spin correlation function given 
by Eq. (\ref{R27}) and $m_1=<s_0>$ is the magnetization of 
the spin in the central site.

Substituting   Eqs. (\ref{R27}) and (\ref{fluc4}) into 
Eq. (\ref{fluc3}), we find 

\begin{eqnarray}
\chi &=& \lim_{n \to \infty} (1-m_1^2)[1+\sum_{j=1}^nq(q-1)^{j-1}{\lambda}^j]
\nonumber \\  
&=&\lim_{n \to \infty}(1-m_1^2)\left\{\frac{1+\lambda}{1-(q-1)\lambda}-q{\lambda}\frac
{[(q-1){\lambda}]^{n-2}}{1-(q-1)\lambda}\right\}.
\label{fluc5}
\end{eqnarray}
It is now clear that the susceptibility $\chi$ diverges for
 $(q-1)\lambda \ge 1$ and given by Eq. (\ref{R36})  
for $(q-1)\lambda < 1$.
>From Eq. (\ref{R36}) we can easily established the 
relation between susceptibility $\chi$ and the correlation 
length $\xi$ in the critical region 
$$
\chi \sim \xi.
$$

\begin{center}
{\bf IV. SPIN - 1 ISING MODEL}
\end{center}

Let us consider, for example, a spin - 1 Ising model, which 
is known as the Blume-Emery-Griffits (BEG) model \cite{beg71}. 
The BEG model on the Bethe lattice was studied in 
Refs. \cite{cm85,aai91}. The model has played an
important role in the development of the theory of tricritical
phenomena \cite{ls84}.

The Hamiltonian of the spin-1 Ising model on the Bethe 
lattice is given by   

\begin{equation}
-\beta H = J\sum_{<ij>}s_is_j + K\sum_{<ij>}s_i^2s_j^2 - 
 \Delta \sum_is_i^2 + h\sum_is_i,
\label{R37}
\end{equation}
where $\beta = (k_BT)^{-1}$ and $s_i = +1, 0, -1$ is the spin 
variable at site $i$. The first term describes the ferromagnetic 
coupling $(J)$ between the spin at sites $i$ and $j$ and the 
second term describes the biquadratic coupling $(K)$. 
Both interactions are restricted to the $q$ nearest neighbor 
pairs of spins. The third term describes the single ion 
anisotropy $\Delta$ and the last term represents the effects 
of an external magnetic field $(h)$.

This model has two order parameters, one is the thermal average 
of the total spin $m_1 = <s_0>$ and the other is the 
quadrupolar moment $m_2 = <s_0^2>$ which reflects the 
possibility of phase separation. These order parameters are 
expressed by 

\begin{equation}
m_1 = \frac{\exp{(h - \Delta)}y^{q} - \exp{(- h - \Delta)}x^{q}}
{1 + \exp{(h - \Delta)}y^{q} + \exp{(- h - \Delta)}x^{q}}
\label{R38}
\end{equation} 
\begin{equation}
m_2 = \frac{\exp{(h - \Delta)}y^{q} + \exp{(- h - \Delta)}x^{q}}
{1 + \exp{(h - \Delta)}y^{q} + \exp{(- h - \Delta)}x^{q}}.
\label{R39}
\end{equation} 
>From Eq. (\ref{R12}) we can obtain the following 
expression for the first - neighbor spin- spin 
correlation function $<s_0s_1>$

\begin{equation}
<s_0s_1>=\frac{(e^{2h}y^{2\gamma}+e^{-2h}x^{2\gamma})e^{(J+K-2\Delta)}-
2x^{\gamma}y^{\gamma}e^{(-J+K-2\Delta)}}
{1+2(e^{h}y^{\gamma}+e^{-h}x^{\gamma})e^{-\Delta}+(e^{2h}y^{2\gamma}+
e^{-2h}x^{2\gamma})e^{(J+K-2\Delta)}-2x^{\gamma}y^{\gamma}e^{(-J+K-2\Delta)}}
\label{cor1}
\end{equation}
where 

$$
x = \lim_{N \to \infty} \frac{g_N(-)}{g_N(0)} \quad  \mbox{and} \quad  
y = \lim_{N \to \infty} \frac{g_N(+)}{g_N(0)}.
$$

Let us introduce the new variables 
$$
v=\frac{x-y}{2} \quad \mbox{and} \quad 
u=\frac{x+y-2}{2},
$$ 
then we obtain 

\begin{equation}
m_1 = -v\frac{u(1+a)+a}{b + u^2+a v^2},
\label{R40}
\end{equation}
 
\begin{equation}
m_2 = \frac{u(u+1)+a v^2}{b + u^2+a v^2}
\label{R41}
\end{equation} 
and
\begin{equation}
<s_0s_1> = \frac{bu^2 + a^3(b + 1)v^2}{ab(b + u^2 + av^2)},
\label{cor2}
\end{equation} 
where $u, v$ are the solution of the recursion relation given 
by Eq. (\ref{R13}) in the thermodynamic limit ($N \to \infty$)

\begin{equation}
\exp{(2h)} = \frac{u - a v}{u + a v}\left(\frac{u+1+v}
{u+1-v}\right)^{q-1} \label{R42},
\end{equation}

\begin{equation}
\exp{(2\Delta)}=\frac{4(b - u)^2}{u^2 - a^2v^2}
\left[(u+1)^2 - v^2\right]^{q-1},
\label{R43}
\end{equation}
and $b$, $a$ are the following constants

$$
b=\exp{(K)}\cosh{(J)}-1,
$$
$$
a=\frac{\exp{(K)}\cosh{(J)}-1}{\exp{(K)}\sinh{(J)}}.
$$

It follows from Eq. (\ref{R19}) that the
spin - spin correlation function for 
spin - 1 Ising model can be written as
\begin{equation}
{\Gamma}(n) = m_1^2 + \sum_{k=1}^{2}A_k\left(\frac{{\lambda}_{k+1}}
{{\lambda}_1}\right)^n 
\label{R44}.
\end{equation}
where $A_1$ and $A_2$ take the form (see Appendix)
$$
A_1=\frac{{\lambda}_3(m_2-m_1^2)-{\lambda}_1(<s_0s_1> - m_1^2)}
{{\lambda}_3 - {\lambda}_2},
$$
and
$$
A_2=\frac{{\lambda}_2(m_2-m_1^2)-{\lambda}_1(<s_0s_1> - m_1^2)}
{{\lambda}_2 - {\lambda}_3},
$$
with $m_1$, $m_2$ and $<s_0s_1>$ given by Eqs. (\ref{R40}), 
(\ref{R41}) and (\ref{cor2}), respectively. The eigenvalues 
${\lambda}_1$, ${\lambda}_2$ ${\lambda}_3$  of the symmetric 
$3{\times}3$ matrix ${\bf V}$  

$$
{\bf V} = \left(\begin{array}{ccc}V_{11}&V_{12}&V_{13}
\\V_{12}&V_{22}&V_{23}\\V_{13}&V_{23}&V_{33} \end{array}\right),
$$
can be obtained from the characteristic equation

$$
(V_{11} - {\lambda})(V_{22} - {\lambda})(V_{33} - {\lambda}) 
+ 2V_{12}V_{13}V_{23}
$$
$$
= V_{23}^2(V_{11} - {\lambda}) + V_{13}^2(V_{22} - {\lambda}) 
 + V_{12}^2(V_{33} - {\lambda}).  
$$
The elements of the matrix $\bf V$ are given by
\begin{equation}
V_{ss'} = \exp{\left(Jss'+ Ks^2s'^2 - \Delta \frac{s^2 + s'^2}{2} 
 + h\frac{s+s'}{2}\right)}
\left[x(s)x(s')\right]^{\frac{\gamma-1}{2}} 
\label{R45},
\end{equation}
where $s$ and $s'$ may independently take values $+1, 0, -1$.

Using Eqs.(\ref{R42}), (\ref{R43})  and (\ref{R45}) we find that 
\begin{equation}
\lambda_1=\frac{b}{b-u},~~
\lambda_{2,3} =\left ( C {\pm} \sqrt{C^2-D} \right ) \lambda_1,
\label{R46}
\end{equation}
where
$$
C=-\frac{u}{2b}+\frac{(ab+a+b)(u^2+u-av^2)}{2ab[(u+1)^2-v^2]},
$$
and 
$$
D=\frac{(b-u)(u^2-a^2v^2)}{ab[(u+1)^2-v^2]}.
$$

The global phase diagram of spin-1 Ising model on the Bethe 
lattice has been studied in detail in the 
Refs. \cite{cm85,aai91}. The ${\Lambda}$ - line of the phase 
transition in the ($J$, $K$ and $\Delta$) space given by 
following conditions:
\begin{equation}
h=0 \quad \mbox{and} \quad \exp{({\Delta}_{\Lambda})} = 
\frac{2(b-u_c)}{u_c}(u_c+1)^{q-1},
\label{R47}
\end{equation}  
where
$$
u_c=\frac{a}{q-1-a}.
$$
In terms of the $T$, ${\Delta}_{\Lambda}/J$ and $K/J$, 
Eq. (\ref{R47}) of the $\Lambda$ - line implies a relation 
$T = T_c({\Delta}_{\Lambda}/J, K/J)$ which locates the 
critical temperature as a function of ${\Delta}_{\Lambda}/J$ 
and $K/J$. 

The critical line starts at ${\Delta}_{\Lambda} \to -\infty$, 
$T_c/J=\frac{1}{2}\ln{\frac{q}{q-2}}$, which correspond the 
critical temperature of the spin - 1/2 Ising model. We note 
that for $\Delta \to -\infty$, the state $s_i=0$ is suppressed, 
and the Hamiltonian (\ref{R37}) reduces to the spin - 1/2 
Ising model with interaction $J$ and external magnetic field 
$h$. For certain values of $K/J$ the system possesses a 
tricritical point at which the phase transition changes from 
the second to the first order. A tricritical point satisfies 
the following equation \cite{aai91}
\begin{equation}
\frac{u_c+1}{\beta-u_c}=q-2+\frac{q-3}{2q}\frac{1}{u_c}.
\label{R48}
\end{equation}
In particular 
$$
\frac{J}{T_{t}}=\frac{1}{2}\ln{\frac{q(3q-2)}{3(q-2)}}, 
\quad \mbox{when} \quad \frac{K}{J}=1,
$$
and
$$
\frac{J}{T_t}=\frac{1}{2}\ln{\frac{-q+6+\sqrt{49q^2-36q+36}}{6(q-2)}}, 
\quad \mbox{when} \quad \frac{K}{J}=3.
$$

Let us now consider the general behavior of the spin-spin 
correlation function in the critical region.
First consider the case $h=0~ (v=0)$, then we have
$$
{\Gamma}(n)=\frac{u(u+1)}{u^2+b}{\lambda}^n
$$ 
with 
$$
{\lambda} = \frac{{\lambda_2}}{{\lambda}_1}=\frac{u}{a(u+1)}.
$$

On the critical line ($T=T_c$, $h=0$ and 
$\Delta = {\Delta}_{\Lambda}$) given by Eq.(\ref{R47}), 
the spin-spin correlation function can be expressed as
\begin{eqnarray}
{\Gamma}(n) &=& \frac{u_c(u_c+1)}{u_c^2+b}\left[\frac{u_c}
{a(u_c+1)}\right]^n
\nonumber \\ 
 &=& \frac{a(q-1)}{a^2+b(q-1-a)}(q-1)^{-n}.
\label{R49}
\end{eqnarray}

By analogy with spin - 1/2 case we can write 
 
\begin{equation}
{\Gamma}(n) = \frac{a(q-1)}{a^2+b(q-1-a)}n^{-d},
\label{R50}
\end{equation}
where $d$ is dimension of the Bethe lattice. 
For $T_c/J=\frac{1}{2}\ln{\frac{q}{q-2}}$ , we obtain 
$\Gamma(n)=(q-1)^{-n}$, which corresponds the case 
of the spin-1/2 Ising model. In the 
critical region, $\Gamma (n)$ has an asymptotic decay of the form

\begin{equation}
{\Gamma}(n) \sim \frac{\exp{\left(-\frac{n}{\xi}\right)}}{n^{d}}
\label{R51},
\end{equation}
where $\xi$ is correlation length and is given by 
$$
\xi = \left[\ln{\frac{1}{(q-1)(\lambda_2/\lambda_1)}}\right]^{-1}.
$$

Near the critical ${\Lambda}$ - line, setting 
$u-u_c=(u_c+1){\delta}$, $v=(u_c+1){\varepsilon}$ and 
$T = T_c(1+t)$, the $h$, $\Delta$ and ${\lambda}_2$ may be 
expanded, for small $\delta$, $\varepsilon$ and $t$ as 
\begin{equation}
h = {\gamma}{\varepsilon}[h_0t-(a-1)\delta + \frac{({\gamma}^2-1)}{3}
{\varepsilon}^2 + (a^2-1){\delta}^2   
- ({\gamma}^2a-1){\varepsilon}^2{\delta} + \frac{{\gamma}^4-1}{5}
{\varepsilon}^4], 
\label{R52}
\end{equation}

\begin{equation}
\Delta - {\Delta}_{\Lambda} = -{\Delta}_0t + (\gamma-a-b)\delta + 
{\gamma}\frac{\gamma-1}{2}{\varepsilon}^2 
 + \frac{a^2-b^2-\gamma}{2}{\delta}^2 -({\gamma}a-1){\gamma}
{\varepsilon}^2{\delta}+{\gamma}\frac{{\gamma}^3-1}{4}{\varepsilon}^4,
\label{R53}
\end{equation}
where $\gamma = q-1, a=(u_c+1)/u_c$, $b=(u_c+1)/(\beta - u_c)$,  
$$
h_0=\frac{J_c(\exp{K_c}-\cosh{J_c})-K_c\sinh{J_c}}{(\exp{K_c}
\cosh{J_c}-1)\sinh{J_c}},
$$
and
$$ {\Delta}_0=\exp{K_c}\frac{J_c\sinh{J_c}+K_c\cosh{J_c}}
{\exp{K_c}\cosh{J_c}-1}.
$$

Consider the case, when $h=0$ and $\Delta={\Delta}_{\Lambda}$. 
>From Eqs. (\ref{R52}) and (\ref{R53}) we obtain
\begin{equation}
c_1t=c_2{\delta}+c_3{\delta}^2
\label{R54}
\end{equation}
with
$$
c_1={\Delta}_0+\frac{3{\gamma}}{2(\gamma+1)}h_0, \quad  c_2=
\gamma-1-\frac{u_c+1}{\beta-u_c}+\frac{1}{u_c}\frac{\gamma-2}{2(\gamma+1)}
$$
and
$$
c_3=\frac{a^2-b^2-\gamma}{2}-\frac{3{\gamma}}{\gamma+1}\frac{1}{u_c}- 
\frac{3{\gamma}({\gamma}^2-{\gamma}-3)}
{4({\gamma}^2-1)(\gamma+1)}\frac{1}{u_c^2}.
$$
It is easy to see from Eqs. (\ref{R48}) and (\ref{R54}) that in all 
points on the $\Lambda$-line $t \sim \delta$, except for 
the tricritical point, where $t \sim {\delta}^2$.

Using Taylor expansion of the expression for the correlation length 
$\xi$ by small $\varepsilon$, $\delta$ and $t$ and  
Eqs. (\ref{R48}) and (\ref{R54}), we find that the 
correlation length $\xi$ increases  as the critical point 
is approached according to $\xi \sim t^{-1}$, with critical 
exponent $\nu = 1$, everywhere on the $\Lambda$-line except 
the tricritical point, where $\xi \sim t^{-1/2}$ with 
tricritical exponent ${\nu}_t = 1/2$.

\begin{center}
{\bf V. SUMMARY AND DISCUSSION}
\end{center}

Let us now briefly summarize our results. 
In this paper we consider the most general spin - $S$ model 
on the Bethe lattice in the external magnetic field 
and use the transfer - matrix approach to 
derive the spin-spin correlation function. The general 
spin - $S$ model inclues
the spin-1/2 simple Ising model and BEG model as special cases. 
>From the exact spin - spin correlation 
functions $\Gamma(n)$ of the Bethe lattic Ising model and BEG 
model in an arbitrary magnetic field $h$ and temperature $T$,
the correlation length ${\xi}$  has been determined analytically. 
In the critical region the correlation length $\xi$ of the simple
Ising model is inversely proportional to the distance 
$\frac{T-T_c}{T_c}$ from the critical point. Such 
singular behavior coincides with the correlation length behavior 
in the one - dimensional chain. We obtain also that near 
the transition point the magnetic susceptibility is 
proportional to the correlation length $\xi$.   

Recently, Gujrati \cite{gu95} showed that in many 
cases the behavior on Bethe or Bethe-like lattices are 
qualitatively correct even when conventional mean-field 
theories fail. By a proper choice of these lattices, it 
is possible to satisfy frustrations, gauge symmetries, 
etc., which are usually lost in  conventional mean-field 
calculations, because of the lack of correlations. Such 
correlations are present on the Bethe-like lattices, and 
in this paper we have given the exact expression for such 
correlations.

It should be noted that we can obtain the proper singular 
behavior of $\xi$ because we have used the proper 
thermodynamic limit ($N \to \infty$) to obtain the 
recursion equations and correlation function, i.e. 
Eqs.(\ref{R13}) and (\ref{R14}) for the most general 
spin - $S$ model on the Bethe lattice and we have 
demonstrated the crucial role of Bethe lattice 
dimensionality in determining critical behavior of 
the correlation length. 
 
In conclusion, it must be remarked that the 
transfer - matrix methods discussed in this paper can be 
extended without difficulty to obtain correlation functions 
with singular correlation length for other spin models, e.g. 
Potts model, multilayer Ising model, Ising model with 
competing nearest-neighbor and next nearest-neighbor 
interactions, etc,  on the Bethe and Bethe-like structures. 
This approach should be applicable for gauge models on 
generalized multi-plaquette hierarchical structures as well.

\begin{center}
{\bf ACKNOWLEDGMENT}
\end{center}

This work was supported by the National Science Council of
the Republic of China (Taiwan) under grant numbers NSC 
86-2112-M-001-001. One of us (N. Sh. I.) thanks the German 
Bundasminsterium fur Forschung and Technologie under grant 
no. 211-5291 YPI and INTAS-96-690 for partial financial 
support. 

\vskip 10 mm

\begin{center}
{\bf APPENDIX}
\end{center}
The coefficient $A_k$ $(k=1, 2,..., 2S)$ can be obtained by 
solving the following system of linear algebraic equations

$$
\left\{\begin{array}{lcl}
g_1=A_1+A_2+...+A_{2S}\\
g_2=A_1l_1+A_2l_2+...+A_{2S}l_{2S}\\
\vdots\\
g_{2S}=A_1l_1^{2S-1}+A_2l_2^{2S-1}+...+A_{2S}l_{2S}^{2S-1},\\
\end{array}\right.
$$
where 
$$
g_k \equiv <s_0s_{k-1}> - m_1^2 \quad {\mbox for} \quad k=1,2,...,2S
$$ 
and $l_k \equiv {\lambda}_{k+1}/{\lambda}_1$.

Let us introduce the elementary Lagrange interpolation 
polynomials
$$
L_i(l)=\prod_{j=1 \atop j \ne i}^{2S}\frac{l-l_j}{l_i - l_j}=
\sum_{j=1}^{2S}a_{ij}l^{j-1}, \quad i=1, 2, ..., 2S,
$$
which satisfy $L_i(l_j)=\delta_{ij}$, where $\delta_{ij}$ 
is the symbol Kronecer.
Now, it is  evident that
$$
A_k=\sum_{j=1}^{2S}a_{kj}g_{j}, \quad \quad k=1, 2, ..., 2S. 
$$ 

The elements of $a_{kj}$ are
$$
a_{kj} = \frac{(-1)^jF_j(k)}{P_k(l_k)},
$$
where $P_k(l)$ is $2S-1$ degree polynomials defined by
$$
P_k(l)=\prod_{i=1 \atop i \ne k
}^{2S}(l-l_i), \quad \quad k=1,2,...,2S
$$
and $F_j(k)$ is the elementary symmetric function in 
$2S-1$ variables $l_1, l_2, ...,l_{k-1},l_{k+1},...,l_{2S}$
$$
\begin{array}{lcl}
F_{2S}=1,\\
F_{2S-1}=l_1+...+l_{k-1}+l_{k+1}+...+l_{2S},\\
\vdots\\
F_1=l_1...l_{k-1}l_{k+1}...l_{2S}.\\
\end{array}
$$
Thus, the coefficients $A_k$ will take the form

$$
A_k=\frac{\sum_{j=1}^{2S}(-1)^jF_j(k)}
{\prod_{i=1}^{2S}(l_k-l_i)},\quad  \quad i \ne k.
$$

{\bf Examples:}

{\bf 1. S=1/2}

\begin{equation}
A_1=g_1=<s_0^2> - m_1^2=1 - m_1^2.
\label{A1}
\end{equation}
For spin - $1/2$ Ising model $<s_0^2> \equiv 1$. 

{\bf 2. S=1}

\begin{eqnarray}
A_1&=&\frac{l_2g_1-g_2}{l_2-l_1}
 \nonumber \\ 
&=&\frac{{\lambda}_3(m_2-m_1^2)-{\lambda}_1(<s_0s_1> - m_1^2)}
{{\lambda}_3 - {\lambda}_2}
\label{A2}
\end{eqnarray}
and

\begin{eqnarray}
A_2&=&\frac{l_1g_1-g_2}{l_1-l_2}
\nonumber \\ 
&=&\frac{{\lambda}_2(m_2-m_1^2)-{\lambda}_1(<s_0s_1> - m_1^2)}
{{\lambda}_2 - {\lambda}_3},
\label{A3}
\end{eqnarray}
where $m_2=<s_0^2>$.

{\bf 3. S=3/2}

$$
<s_0s_n> - m_1^2=A_1l_1^n + A_2l_2^n + A_3l_3^n,
$$
with
$$
A_1=\frac{l_2l_3g_1-(l_2+l_3)g_2+g_3}{(l_2-l_1)(l_3-l_1)}
$$
and
$$
A_2=A_1(l_1 \Leftrightarrow l_2), \quad \quad A_3=A_1
(l_1 \Leftrightarrow l_3),
$$
where $g_1 = <s_0^2> - m_1^2, \quad g_2 = <s_0s_1> - m_1^2$ and 
$g_3 = <s_0s_2> - m_1^2$.

{\bf 4. S=2}

$$
<s_0s_n> - m_1^2=A_1l_1^n + A_2l_2^n + A_3l_3^n + A_4l_4^n.
$$

$$
A_1=\frac{F_1g_1-F_2g_2+F_3g_3-F_4g_4}{(l_2-l_1)(l_3-l_1)(l_4-l_1)}
$$
with
$$
F_1=l_2l_3l_4, \; F_2=l_2l_3 + l_2l_4 + l_3l_4, \; 
F_3=l_2+l_3+l_4, \; F_4=1
$$
and
$$
A_k=A_1(l_1 \Leftrightarrow l_k), \quad \quad {\mbox for} \quad k=2,3,4,
$$
where $g_1=<s_0^2> - m_1^2, \quad g_2=<s_0s_1> - m_1^2, 
\quad g_3=<s_0s_2> - m_1^2$ and $g_4=<s_0s_3> - m_1^2$.

\end{document}